\documentclass[aip, amsmath, amssymb, reprint]{revtex4-2}
\usepackage[colorlinks,citecolor=blue]{hyperref}
\usepackage{graphicx}
\usepackage[utf8]{inputenc}
\usepackage{siunitx}

\begin{document}
\title{Direct formation of nitrogen-vacancy centers in nitrogen doped diamond along the trajectories of swift heavy ions}

\author{Russell E. Lake} 
\affiliation{Accelerator Technology \& Applied Physics, Lawrence Berkeley National Laboratory, 1 Cyclotron Road, Berkeley, CA 94720, USA}
\affiliation{QCD Labs, COMP Centre of Excellence, Department of Applied Physics, Aalto University, P.O. Box 13500, FI-00076 Aalto, Finland}

\author{Arun Persaud}  
\email[Author to whom correspondence should be addressed:]{apersaud@lbl.gov}
\author{Casey Christian}  
\affiliation{Accelerator Technology \& Applied Physics, Lawrence Berkeley National Laboratory, 1 Cyclotron Road, Berkeley, CA 94720, USA}

\author{Edward S. Barnard} 
\author{Emory M. Chan} 
\affiliation{The Molecular Foundry, Lawrence Berkeley National Laboratory, 1 Cyclotron Road, Berkeley, CA 94720, USA}

\author{Andrew A. Bettiol} 
\affiliation{Department of Physics, National University of Singapore, 2 Science Drive 3, Singapore 117551}

\author{Marilena Tomut} 
\affiliation{GSI Helmholtz Center for Heavy Ion Research, 64291 Darmstadt, Germany}
\affiliation{Institute of Materials Physics, WWU Münster, Wilhelm-Klemm-Straße 10, 48149 Münster, Germany}

\author{Christina Trautmann} 
\affiliation{GSI Helmholtz Center for Heavy Ion Research, 64291 Darmstadt, Germany}
\affiliation{Technical University of Darmstadt, 64287 Darmstadt, Germany}

\author{Thomas Schenkel} 
\affiliation{Accelerator Technology \& Applied Physics, Lawrence Berkeley National Laboratory, 1 Cyclotron Road, Berkeley, CA 94720, USA}

\date{\today}

\begin{abstract}
    We report depth-resolved photoluminescence measurements of nitrogen-vacancy (NV$^-$) centers formed along the tracks of swift heavy ions (SHIs) in type Ib synthetic single crystal diamonds that had been doped with \SI{100} {ppm} nitrogen during crystal growth.  Analysis of the spectra shows that NV$^-$ centers are formed preferentially within regions where electronic stopping processes dominate and not at the end of the ion range where elastic collisions lead to formation of vacancies and defects. Thermal annealing further increases NV yields after irradiation with SHIs preferentially in regions with high vacancy densities.  NV centers formed along the tracks of single swift heavy ions can be isolated with lift-out techniques for explorations of color center qubits in quasi-1D registers with an average qubit spacing of a few nanometers and of order 100 color centers per micrometer along 10 to 30 micrometer long percolation chains.  
\end{abstract}

\maketitle

The negatively charged nitrogen-vacancy (NV$^-$) defect center in diamond has gained much attention for applications in quantum optics due to its long coherence time at room temperature and its great potential for quantum sensing and quantum communication.\cite{Schroder2016-vz, Wehner2018-sy, Dolde2013-ny} Reliable formation of color centers with long spin coherence times and the placement of color centers into quantum registers are major challenges for applications.  This is, in part, due to the fact that our understanding of the microscopic formation mechanisms of NV centers has remained incomplete. In particular, density functional theory apparently contradicts some experimental results.\cite{Deak2014-yi}  Studies of color center formation processes under a series of experimental conditions can help support advances towards deterministic formation of high quality color centers.\cite{Schroder2016-vz}

From the perspective of quantum device fabrication, it will be useful to place single color centers into precise locations of a sample.  This calls for modes of fabrication with high spatial resolution. Here, we report on the formation of color centers in diamond along the trajectories of swift heavy ions (SHIs).  SHIs, such as gold ions at \SI{1.1}{\GeV}, transverse solids along near straight lines over distances of tens of micrometers with high probability. We report that NV centers form in the interaction of SHIs with nitrogen doped diamond directly, without any thermal annealing.  In many earlier studies, energetic ions and electrons have been used to first form vacancies and then NV-centers in diamonds during a consecutive thermal annealing step.\cite{Meijer2005-297} Hence, we observe an efficient one-step process for local color center formation with alignment of color centers along the trajectory of each SHI.  In an earlier study, we had observed the formation of NV centers following irradiation of diamonds with SHIs where the diamonds had been implanted with nitrogen ions over a depth of about \SI{100}{\nano\meter} near the diamond surface.\cite{Schwartz2014-lc} The motivation for our current study was twofold. In addition to the development of new fabrication tools for diamond-based quantum-photonic devices,\cite{Toyli2010-kg} we also seek to understand the role of intense electronic excitations in color center formation within the nonequilibrium process of track formation.\cite{Koenig_2005}  Swift heavy ion tracks in diamond have recently been treated with computational methods.\cite{Schwen2012-zq,Schwen2007-ne,Valencia2015-aa} Efforts to engineer non-Poissonian spatial distributions of color centers in diamond have included, among others, the use of He ions,\cite{Waldermann2007-pe,Huang2013-xv,Acosta2009-ea} \SI{60}{keV} N ions,\cite{Varichenko1986-he} highly charged Ar ions,\cite{Racke2020-oy} and beams of energetic electrons in preimplanted diamond.\cite{Schwartz2012-ma} NV center yields of up to about 50\% have been reported,\cite{Pezzagna2010-db} but much lower yields are common.\cite{botsoa-prb-2011} Recently, progress has been reported on achieving a higher NV center yield using ion implantation and local doping of diamonds.\cite{Luhmann2019-se} While higher color center yields have been reported, underlying mechanisms of competing color center processes remain to be disentangled and formation of spin-photon qubit registers with (sub-)\SI{10}{\nano\meter} resolution as required for nearest neighbor coupled qubits with magnetic dipolar interaction remains a challenge. 

\begin{figure}[htb]
\includegraphics[width=\linewidth]{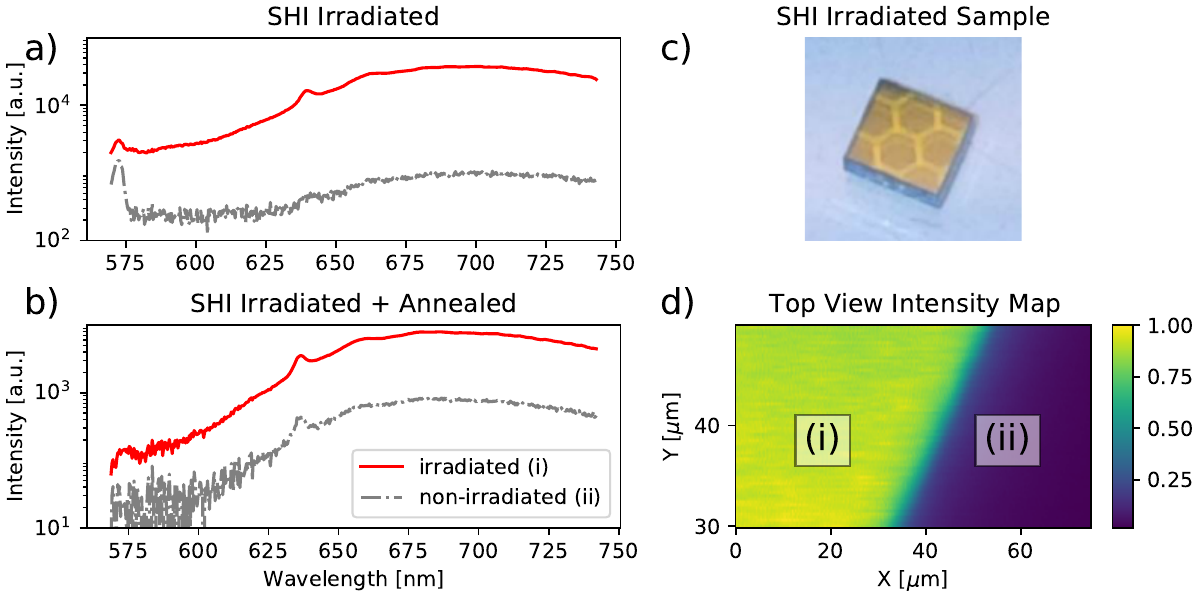}
\caption{Photoluminescence spectra of SHI-irradiated diamond. The plots a) and b) show spectra from the SHI-irradiated and later annealed samples.  A photograph of the sample is shown in c) where the lighter regions are pristine and dark regions are irradiated.  In d), a normalized count rate intensity map at fixed depth shows contrast between irradiated (i) and pristine (ii) regions of the irradiated and annealed sample. Note: the absolute values in a) cannot be directly compared to the results in b).\label{fig:spectra}}%
\end{figure}

The samples studied in this work where type Ib diamonds (Element Six) with approximately \SI{100}{ppm} nitrogen present from chemical vapor deposition growth. During irradiation with swift heavy ions, the sample was masked by a thick metallic grid with millimeter-sized openings to allow for both irradiated and nonirradiated regions to be measured at the same time.  Irradiations were carried out at the Helmholtzzentrum für Schwerionenforschung (GSI) with \SI{1.1}{\GeV} $^{197}$Au ions. The irradiation was performed in high vacuum and nominal at room temperature with the sample mounted on an aluminum holder that also acted as a heat sink. Beam-induced macroscopic heating of the sample is estimated to stay well below \SI{80}{\celsius}, based on temperature monitoring using an infrared camera of similar sized carbon samples irradiated under the same beam conditions. The diamond was irradiated with a fluence of \SI{1e12}{ions \per \centi \meter \squared} using an ion beam flux of \SI{6e9}{ions \per \centi \meter \squared \per \second}. Gold ions at these energies have a range of \SI{34}{\micro \meter} in diamond.  After irradiation, the mask was removed and contrast between the irradiated and nonirradiated regions is visible by eye (Fig.~\ref{fig:spectra}c), where the darker regions of the sample were irradiated and the lighter regions were masked. 

\begin{figure}[htb]
\includegraphics[width=\linewidth]{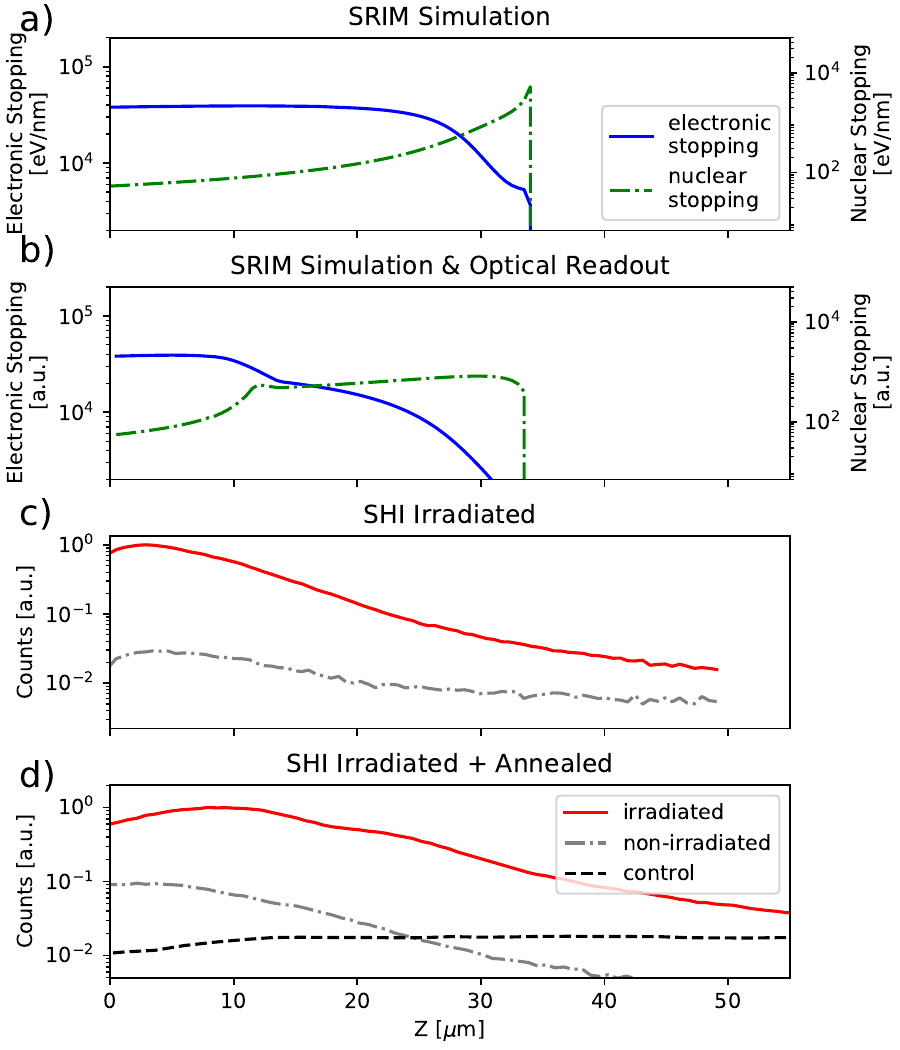}
\caption{a) SRIM simulation of \SI{1.1}{\GeV} Au ions into diamond showing contributions of the electronic and nuclear stopping processes along the ion paths. b) SRIM data convoluted with depth-depended probe volume. c) Depth profiles of the NV$^-$ peak for the SHI-irradiated sample. d) Depth profiles of the NV$^-$ peak after annealing the sample combined with a measurement from a control sample with a uniform NV$^-$ distribution. Note: the absolute values in d) cannot be directly compared to the results in c).\label{fig:depth}}
\end{figure}

We performed room temperature photoluminescence (PL) measurements across boundaries of the masked and irradiated regions. These measurements were performed with a custom built confocal PL setup designed for spatially resolved three dimensional maps of optically active defects in semiconductors with a similar experimental setup to Ref.~\onlinecite{Barnard2013-lr}.  The sample was mounted with adhesive tape on a glass slide and placed on a piezo-nanopositioning sample stage.  Control experiments on a reference sample (Element Six, with a nitrogen concentration of \SI{1}{ppm}) demonstrated that using this apparatus we can collect PL signals from the NV$^-$ defects at depths of \SI{>60}{\micro\meter}, see Fig.~\ref{fig:depth}d.  Using the depth-resolved one-photon excitation-collection scheme,\cite{Barnard2013-lr} photoluminescence was collected through a pinhole and into detection optics. For spectral measurements, we used an Acton 2300i spectrometer with 150 groves/mm grating and an Andor iXon electron-multiplied CCD. The excitation wavelength was \SI{532}{\nano\meter}. 

\begin{figure}[htb]
\includegraphics[width=0.65\linewidth]{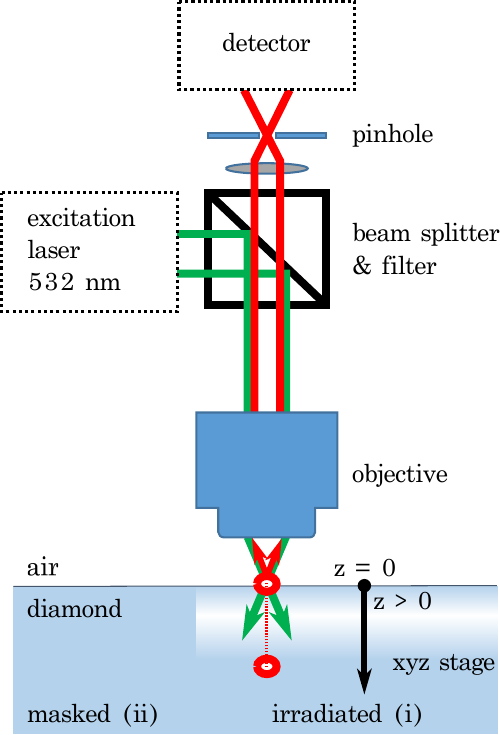}
\caption{Experimental schematic for three-dimensional photoluminescence measurements. \label{fig:schematic}}%
\end{figure}

Initial measurements were performed after irradiating, and before annealing the sample, and are shown in Fig.~\ref{fig:spectra}a. PL spectra are measured within the irradiated (red line) and masked (gray line) regions.  The intensity of the PL is enhanced in the regions that were irradiated by the ions. Qualitatively similar spectra are observed after annealing the sample in high vacuum at \SI{800}{\celsius} for one hour in Fig.~\ref{fig:spectra}b. The difference in PL intensity between irradiated and nonirradiated areas can also be clearly seen in the contrast of Fig.~\ref{fig:spectra}d, which shows a $x$-$y$ map of the sample and the color represents the integrated measured PL spectra (normalized). 
The intensity of the PL spectra is measured in units of total photon counts per second within the wavelength span from \SIrange{570}{740}{\nano\meter} and is at least a factor of ten higher in the irradiated region compared to the nonirradiated region.
We track the depth dependence of the spectral feature at $\lambda = \SI{637}{\nano\meter}$ associated with the zero phonon line (ZPL) of the NV$^-$ centers using the experimental setup shown schematically in Fig~\ref{fig:schematic}.
In particular, we move the $x$-$y$-$z$ sample stage while keeping the excitation laser focus position constant.  We can thus scan the sample stage position to collect PL signal from a selected volume within the sample.

To locate the stage position where the laser focuses on the surface, we add a filter to remove the PL contribution and record only the $z$-dependent reflection signal. The stage position at maximum reflection is identified as the surface position $z = 0$ and used to calibrate the $z$ axis in Fig.~\ref{fig:depth}. 

As the stage moves toward the objective and the focus spot moves deeper into the sample, we record the PL spectrum for each location for \SI{100}{\milli \second}.  To convert from the stage position coordinate to the actual depth where the diamond is optically excited, we multiply the stage position by the index of refraction in diamond, $n=2.4$, as $z=n \, z_{\text{stage}}$. Although the stage has nanometer resolution, the depth resolution is given by light excitation and collection volumes, which are stretched due to the high index of refraction of diamond. Using Eq.~1 from Ref.~\onlinecite{Everall2000-gy}, we find that in our measurements, where we used a 100$\times$ lens with a numerical aperture (NA) of 0.95, the resolution scales proportionally to $4.6 z_{\text{stage}}$ as we excite deeper in the sample. 

To analyze the data, we average the spatial regions to achieve better signal-to-noise spectra for the irradiated versus nonirradiated region (Fig.~\ref{fig:spectra}) or integrate just the NV peak at $\lambda=\SI{637}{\nano\meter}$ from \SIrange{630}{645}{\nano\meter} and the lateral spacial dimension to obtain the depth distribution of the NV centers as shown in Fig.~\ref{fig:depth}.

Figure~\ref{fig:depth} displays the depth profile of the integrated PL around the NV$^-$ ZPL line for the areas irradiated to SHI radiation (red) and the nonirradiated control regions (gray). Data from each panel are normalized to the maximum intensity measured in the irradiated region. In Figure~\ref{fig:depth}a, the results from Stopping Ranges in Matter (SRIM)\cite{Ziegler2010} are shown. In Figure~\ref{fig:depth}b, we assume that NV centers are produced exclusively by either electronic or nuclear stopping and calculate the expected response by folding the simulated data from panel a) with the estimated probe volume. For the SHI-irradiated sample in Fig.~\ref{fig:depth}c, we observe an increase in intensity within the irradiated region between the surface and a depth of approximately \SI{5}{\micro\meter} followed by a drop in intensity. The nonirradiated region of the sample shows a much weaker signal with a similar profile (possible due to light scattering). After the sample was annealed, we observe a similar characteristic profile with an additional shoulder at a larger depth (\SIrange[range-units = single, range-phrase=--]{20}{35}{\micro\meter}). The anneal also activates NV center in the nonirradiated regions although at a much lower intensity level.  The measurements in Figure~\ref{fig:spectra}a/b and Figure~\ref{fig:depth}c/d were taken at different times and due to parameter changes in the confocal microscope, one cannot directly compare absolute numbers between these two figures (irradiated and nonirradiated data in each subplot can be compared though). Furthermore, we want to mention that the purpose of the control sample shown in Figure~\ref{fig:depth}d is to verify that we can acquire a NV signal below the end-of-range depth of the SHI, but the amplitudes cannot be compared to the data from the irradiated diamond sample, because of unknown NV concentration in the samples.

Using a diamond with a uniform distribution of nitrogen, we observe that NV centers are created by SHIs. The observed depth profile agrees well with the profile of the electronic stopping once we take the changes in probed volume into account. We also observe no peak in the NV center density that can be contributed directly to the end-of-range peak of nuclear stopping of the SHIs in diamond (simulated using SRIM\cite{Ziegler2010}). We attribute the difference between the irradiated and nonirradiated signal in Figure~\ref{fig:depth}c at a depth larger than the range of the SHI to light collection from out of focus areas, e.g., from NV centers at a shallower depth. We also note that a standard anneal of the sample only creates a small fraction of the NV centers the SHIs created.  After annealing the sample, we observe additional NV centers at the end-of-range of the ions where most of the vacancies are created. Therefore, we conclude that although vacancies are needed for the NV formation, the availability of vacancies alone is not enough to create NV centers and that energy deposition due to electronic stopping or ion recoils also plays an important role in effectively creating NV centers. 

We estimate the self-absorption of the emitted light by the NV centers by using an absorption cross section\cite{Subedi2019-xe} of $\sigma=\SI{2.8e-17}{\cm \squared}$ and a density of $d=\SI{1.76e19}{\per\cm\cubed}$ (equivalent to \SI{100}{ppm} and \SI{100}{\percent} activation) and calculating the transmission $T=\exp\left(-l \sigma d \right)$, where $l$ is the length of the photon path. This results in a transmission larger then \SI{98}{\percent} at a depth of \SI{40}{\micro\meter}. We therefore can ignore the effect of self-absorption.

The observed NV$^-$ distribution does seem to follow the electronic energy loss of the SHI closely. A possible explanations could be that this is due to the role of secondary electrons (or delta-electrons) that can be formed in close collisions of SHI with target electrons.  Electrons with energies up to \SI{10}{\keV} can be formed in collisions with \SI{1}{\GeV} gold ions, well below the threshold for vacancy formation of \SI{120}{\keV}.\cite{Schwartz2012-ma}  Electrons can contribute in several ways to creating NV centers, for example, by heating and locally annealing the volume around the SHI track.  Detailed energy-loss and energy transport simulations\cite{Barnard2017-jh} and further experiments are ongoing. Another reason for a decrease of the NV$^-$ signal near the end of range of the gold ions could be the formation of other defects, for example, vacancy clusters, which could lower the PL signal.

The primary effect of thermal annealing is the activation of NV centers due to the initial nitrogen concentration in the samples (irradiated and nonirradiated areas). Furthermore, in the SHI-irradiated regions, the annealing causes a broadening of the intensity-depth distribution due to diffusion and an increased activation near the end-of-range of the ions that we attribute to the additional vacancies that are available in this region (\SIrange[range-units = single, range-phrase=--]{20}{35}{\micro\meter}). By considering the difference in intensity between the irradiated and nonirradiated spectra of Fig.~\ref{fig:depth}d, we observe that the \SI{800}{\degreeCelsius} anneal does not reproduce the effect of SHI radiation exposure. Specifically, the intensity of the NV$^-$ peak is approximately 50 times higher in the irradiated region than in the non-irradiated-annealed region before annealing. Measurements of quantitative changes in NV$^-$ population due to ion irradiation (with and without annealing) are still in progress. 

From the increase of the SHI-induced NV$^-$ intensities compared to nonirradiated regions in our annealed diamond, we can estimate a NV$^-$ conversion efficiency of SHIs. First, we observe that the intensity of the irradiated region is roughly 8$\times$ larger than in the nonirradiated region. Assuming a nitrogen-to-NV$^-$ conversion factor, $p$, during the anneal, a depth-independent conversion factor, $q$, from the SHI, and a total number of available nitrogen, $A$, we can calculate the number of NV$^-$ centers in the nonirradiated, annealed region to be $Ap$ and in the irradiated region to be $Aq + A(1-q)p$ (assuming we do not lose NV$^-$ centers from SHIs during the anneal). The ratio, $r$, of the SHI conversion factor and the annealing conversion factor can then be expressed as $r=\frac{q+(1-q)p}{p}$. This results in a SHI conversion factor of $q=(r-1)\frac{p}{1-p}$. Rabeau \textit{et al.} report a conversion factor using a similar anneal as we do of $p=\SI{2.5}{\percent}$,\cite{Rabeau2006-ec7} which would result in a conversion factor for SHI between 15\% and 20\%. However, this is only a rough estimate of NV$^-$ formation yield from SHI, ignoring, for example, its depth dependence. Future experiments will include the measurement of the activation by annealing ($p$) and improved depth-measurements allowing a better estimate of the NV$^-$ conversion factor for SHIs.
 
SHIs transverse diamond samples on almost straight lines, with effective interaction track diameters of a few nanometers radius\cite{Schwen2007-ne} and trajectory lengths of tens of micrometers.  NV centers, hence, form along a quasi-1D chain.  The apparent density of NV centers in SHI-irradiated areas is about 50$\times$ higher than in nonirradiated regions. Their average spacing will be given by the concentration and distribution of the nitrogen (present mostly as P1 centers) following single crystal growth together with the NV-formation efficiency per SHI. Thermal annealing will broaden the distribution due to defect diffusion processes.\cite{Acosta2009-ea} 1D chains of NV centers from low fluence irradiations (e.g., \SI{1e6}{ions \per \centi \meter \squared}) can be isolated from bulk samples using common lift-out techniques see, e.g., Chai \textit{et al.},\cite{Chai2007-gv} and they can then be integrated, for example, with microwave sources and magnetic fields for exploration of spin-photon qubits with nearest neighbor coupling along a percolation chain. We plan to perform lift-out of NV chains created by SHI irradiation in future experiments.  Using the estimated NV conversion efficiency of 15\%-20\% and the N density of \SI{100}{ppm}, \SI{1.1}{\GeV} gold ions form NV centers with an average spacing of a few nanometers over a distance of over ten micrometers in a quasi-1D register with potentially over one thousand qubits.  A high nitrogen background concentration stabilizes NV-center charge states, but the nitrogen spin bath will also limit spin coherence times to a few microseconds at room temperature.\cite{Van_Wyk1997-rc}

The work at LBL was supported by the Office of Science, Office of Fusion Energy Sciences, of the U.S. Department of Energy, and Laboratory Directed Research and Development (LDRD) funding from Berkeley Lab, provided by the Director, Office of Science, of the U.S. Department of Energy; and the work at the Molecular Foundry was supported by the Office of Science, Office of Basic Energy Sciences, of the U.S. Department of Energy, under Contract No. DE-AC02-05CH11231.
The ion irradiation at GSI are based on a UMAT experiment, which was performed at the M-branch of the UNILAC at the GSI Helmholtzzentrum für Schwerionenforschung, Darmstadt (Germany) in the frame of FAIR Phase-0.
R.E.L. acknowledges support from the Academy of Finland under grant No.~265675 and Aalto University Centre for Quantum Engineering.
M.T. acknowledges funding from the European Union’s Horizon 2020 Research and Innovation program under Grant Agreement No 730871.
This work was also supported by the coordinated research project ``F11020'' of the International Atomic
Energy Agency (IAEA). 

\section*{Data Availability}

The data, analysis scripts, and simulation scripts are openly available on Zenodo at \url{https://doi.org/10.5281/zenodo.4018144}, Ref~\onlinecite{data}.

\section*{References}

\bibliography{russell}

\end{document}